\documentclass[sigplan,10pt]{acmart}

\renewcommand\footnotetextcopyrightpermission[1]{}
\usepackage{prp-macros}
\renewcommand{\tinyskip}{\vspace{1pt}}
\usepackage{framed}

\newcommand{\sys}{\textsc{Planarian}\xspace}
\newcommand{\vanilla}{\textsc{No-snapshot}\xspace}
\newcommand{\export}{\textsc{Full-snapshot}\xspace}

\usepackage[small, compact]{titlesec}

\begin{document}

\title{\sys{}: Managing Agent State with Statepoints}

\author{Jinnan Guo}
\affiliation{%
 \institution{Imperial College London}
 \country{}
}

\author{Hao (Mark) Chen}
\affiliation{%
 \institution{Imperial College London}
 \country{}
}

\author{Kapil Vaswani}
\affiliation{%
 \institution{Indian Institute of Science Bangalore}
 \institution{SPARC}
 \country{}
}

\author{Andrew Paverd}
\affiliation{%
 \institution{Microsoft Security Response Center}
 \country{}
}

\author{Peter Pietzuch}
\affiliation{%
 \institution{Imperial College London}
 \country{}
}

\begin{abstract}

LLM agents solve complex tasks by iteratively changing files, invoking local tools, and interacting with remote services, which modifies state across their local environment and remote services. Today, agents and users must manage these changes explicitly, whether reverting exploratory actions or recovering from erroneous ones. Doing so safely requires coordinated actions, yet current agent harnesses lack unified abstractions and mechanisms for managing local and remote state consistently and efficiently.

We describe \sys{}, an agent runtime with state management that enables agents and users to recover from erroneous actions and explore alternative executions over consistent local and remote environment state. \sys introduces the abstraction of \emph{agent statepoints}, which are consistent, restorable point-in-time versions of the environment state. \sys exposes three state-management primitives to agents and users: (i)~\emph{snapshot} creates a new statepoint spanning local and remote state without requiring external services to support checkpoints: it relies on efficient incremental process and file system snapshotting to capture local sandboxed state, and transparently records compensating actions to undo remote state changes; (ii)~\emph{rollback} restores the environment to a previous statepoint by reverting to a prior local checkpoint and replaying compensating actions for remote state changes; and (iii)~\emph{fork} creates multiple isolated branches from a statepoint, enabling the agent to explore alternatives in parallel. We show that \sys enables agents to undo mistakes and explore alternatives in parallel, improving task quality by up to 15$\times$, and allows users to recover from erroneous actions with only 3\% overhead.


\end{abstract}

\maketitle

\pagestyle{plain}


\section{Introduction}
\label{sec:intro}

Agents perform tasks that require repeated interaction with a host machine, from modifying software to administering services~\cite{claudecode, codex, openclaw}. An \emph{agent harness} invokes a large language model~(LLM) with \emph{context} comprising instructions, prior actions, and tool results. It uses the model's output to invoke \emph{tools} and incorporates their results into subsequent context. Tool calls request commands or service operations that act on \emph{environment state}: local files and processes, and state held by external services, such as database state. Local tools can execute directly on the host or within a \emph{sandbox} that isolates their execution. The harness can invoke remote tools through the \emph{model context protocol}~(MCP)~\cite{mcp}.

When an agent fails to complete a task, it may have to undo earlier actions before trying different steps. A user may also want to reverse an unwanted change, such as the deletion of files or incorrect updates to a remote calendar service. For the agent, reconstructing state requires extra tool calls, which may be costly or impossible if data was lost. We observe that explicit \emph{state management} would let agents and users recover from mistakes without manually having to repair the affected state. Agents could also explore the effect of alternative actions on the environment efficiently without repeating successful work or carrying over unwanted effects.

Supporting such state management alongside existing agent harnesses, however, raises three challenges:

\mypar{(1)~Consistency across local and remote state} State management requires \emph{environment consistency}, \ie it must restore both local and remote state consistently. Local files and processes can be checkpointed, but MCP endpoints of remote network services may provide neither snapshots nor undo operations. Recovering only local state can result in inconsistencies though: for example, a data-cleaning agent may update database records through an MCP endpoint and write a local report of its changes. Reverting only the local report leaves it inconsistent with the updated remote database.

\mypar{(2)~Integration with agent execution} Users and agents must be able to perform state management through existing agent harnesses. In addition, state management must guarantee \emph{context consistency}, \ie the agent must be aware of the changes to the environment state. After the environment state is restored, the history of tool calls may describe effects that no longer exist. Simply retaining this history can mislead the agent; discarding it loses evidence of unsuccessful actions that could guide future actions. State management therefore affects both how the harness dispatches tools and what context it supplies. 

\mypar{(3)~Overhead of state management} Frequent state management must not dominate task execution time or storage use. Local state, including files and process memory, may be large, even when individual tool calls change little state. State management overheads may accumulate both as execution advances and whenever an earlier state is revisited to try different actions. Managing remote state also incurs communication latency and is constrained by MCP interfaces.

\tinyskip

\noindent
Prior approaches solve only some of these challenges. Checkpointing systems such as DMTCP~\cite{dmtcp_ipdps_2009} and Catalyzer~\cite{catalyzer_asplos_2020} preserve local execution state, but cannot account for remote service state or agent decisions. Transactional workflows~\cite{beldi_osdi_2020}, speculative configuration repair~\cite{autobash_sosp_2007}, and intrusion recovery with external compensation~\cite{retro_osdi_2010} establish broader recovery models, but target generic workflows or application repair rather than agent exploration. Without environment and context consistency, state management cannot serve the parties that need it: an agent cannot safely roll back on its own, a harness cannot search over alternative branches, and a user cannot reliably undo an unwanted change. The open problem is thus to make an agent's evolving environment state a unit of recovery and exploration, which is consistent across local and remote state, consistent with the agent's context, and efficient enough to be used repeatedly.

We describe \sys{}, an agent runtime for state management that operates together with existing agent harnesses. \sys provides a unified abstraction for managing local and remote environment state, enabling consistent recovery and exploration while supplying the information needed for subsequent agent decisions. For this, \sys makes three technical contributions:

\mypar{(1)~Agent statepoints and primitives} Recovery and exploration need a common abstraction for managing both local and remote state. \sys introduces \emph{agent statepoints}: a statepoint captures both local files and running processes together with the remote state changed through MCP operations. A statepoint remains restorable even when the remote service cannot take checkpoints, and carries a description of the captured state and the outcomes of prior executions. For state management, three primitives operate over statepoints: (i)~\emph{snapshot} records a new statepoint; (ii)~\emph{rollback} returns to a statepoint to undo state changes; and (iii)~\emph{fork} starts an isolated alternative execution in a branch from a statepoint, while preserving the existing state.

\mypar{(2)~Integrating agent statepoints with agent execution} \sys exposes its state-management primitives (a)~as tools that agents invoke directly, or through APIs for (b)~harness-driven search or (c)~manual user requests. During execution, the agent must select suitable statepoints and interpret them correctly. \sys uses the evidence associated with statepoints both to inform this choice and to update the agent context afterward. The agent harness retains responsibility for LLM invocations and context construction, routing tool calls through \sys and incorporating the returned results. \sys manages environment state through local tool sandboxing and MCP proxying, and captures statepoints between tool calls. After restoration, \sys supplies the selected state's description and prior outcomes to the agent harness to append to the context. This identifies the environment state now available to agents without discarding the information needed to choose different actions.

\mypar{(3)~Local checkpointing and remote compensation} To keep the overhead of state management low, \sys limits the state that must be copied locally and reconstructed remotely when creating or reverting to statepoints. For \emph{local state}, \sys combines incremental copy-on-write filesystem snapshots~\cite{zfs} with CRIU process checkpoints~\cite{criu} that record only changed memory. \sys re-establishes the restored checkpoint as the baseline for subsequent memory changes, allowing incremental capture across rollback and fork without modifying CRIU or the kernel. This avoids repeatedly copying unchanged memory during exploration. For \emph{remote state}, \sys proxies supported MCP requests so that their effects can be reversed through \emph{compensating actions}---service operations that undo earlier changes. For example, a database MCP proxy preserves original database row values and affected keys within the transaction that modifies them, allowing destructive updates to be undone.

\tinyskip

\noindent
We evaluate \sys with agents in coding, personal assistance, system administration, database management, and gaming tasks. In the Mario platform game, an agent that uses \sys to retry from statepoints achieves 15$\times$ the game score of an agent without state management. When solving system-administration tasks with MCTS-style tree search, \sys takes snapshots 10$\times$ faster than a baseline that uses full process and filesystem checkpoints. Across system and database administration tasks, state management adds only 1\%--3\% to completion time, including coordinated recovery for an agent that modifies a remote database through MCP. These results show that \sys enables agents to revisit decisions and explore alternatives while incurring little overhead for state management.



\section{State Management for Agents}
\label{sec:background}

First we describe agent execution~(\S\ref{sec:background:harness}), establish the requirements for managing environment state~(\S\ref{sec:background:state}), and assess existing state-management approaches~(\S\ref{sec:background:challenege}).

\begin{figure}[t]
  \centering
  \includegraphics[width=0.379130\textwidth]{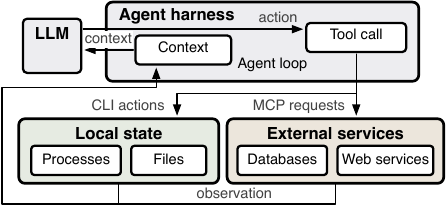}
  \caption{Agent harness invoking the LLM and tools}
  \label{fig:harness}
\end{figure}

\subsection{Agent execution}
\label{sec:background:harness}

As LLMs become capable of using tools, users delegate everyday tasks, such as coding, system administration and data maintenance, to agents for automation~\cite{terminalbench, devopsgym}.

As shown in \F\ref{fig:harness}, an \emph{agent} uses an LLM to choose actions toward a task objective. An \emph{agent harness} coordinates the model and tool invocations through which the agent interacts with this environment through an agent loop~\cite{react}: it invokes the LLM with context, interprets the output to invoke tools, and incorporates tool results into the context for subsequent LLM invocations. The agent's \emph{environment state} thus comprises local machine state, such as files and running processes, and external network service state. 

Agent harness implementations range from general ones (\eg LangChain~\cite{langchain} and AutoGen~\cite{autogen}) to specialized harnesses for coding agents (\eg Claude Code~\cite{claudecode}, Codex~\cite{codex}) and personal assistants (\eg OpenClaw~\cite{openclaw}, Nanobot~\cite{nanobot}).

Tools provide access to local commands and files and remote service operations. An agent harness can invoke local tools on the host and remote tools through the model context protocol~(MCP)~\cite{mcp}. MCP provides a common interface for invoking APIs exposed by remote servers.

Sandboxing~\cite{sandboxing} is an optional safety feature that confines local tool execution to an isolated environment. Restricting access to host files and processes limits the damage that erroneous commands can cause outside the sandbox. Sandboxes can be realized using containers~\cite{linuxcontainers}, which isolate processes and their filesystem view through OS mechanisms, such as Linux namespaces. A harness creates and starts a container from an image containing the required software, executes local tool calls inside it, and removes it when the task ends. Sandboxing libraries and SDKs, such as E2B~\cite{e2b}, provide APIs through which harnesses can use sandboxing.

\subsection{Managing environment state}
\label{sec:background:state}

As an agent executes, it mutates the environment state, producing visible side effects on the machine. When an agent updates environment state, there are two broad requirements:

\mypar{Effectiveness} Agents often solve challenging tasks through search and exploration~\cite{exact}. By sampling multiple rollouts from an LLM, an agent can improve its problem-solving capability by allowing it to try alternative plans and learn from unsuccessful attempts. Such exploration mechanisms are increasingly common: pass@$K$ allows an agent to attempt a task $K$~times~\cite{pass_at_k}; Monte Carlo tree search~(MCTS)~\cite{mcts} performs exploration and backtracking as part of the harness; and EAPO~\cite{eapo} encourages the agent to explore and recover from mistakes through backtracking.

At the system level, an agent must be able to explore plans sequentially or concurrently, with some inevitably failing. These approaches impact the environment state, as they require backtracking and state branching so that side effects of old plans do not affect subsequent execution.

\mypar{Safety} Since LLM outputs are non-deterministic and unreliable (\cf hallucination~\cite{hallucination}, sycophancy~\cite{syceval}), an agent's tool calls can produce unintended, harmful side effects, \eg deleting required files, installing incorrect packages, or committing incorrect writes to a database. Unlike a conventional program, an agent writes and executes commands without human review, so it may break a production environment before the mistake is noticed by the user. Running an agent safely thus requires the ability to \emph{rollback} the state after a mistake, preserving the integrity of the environment state.

\begin{table*}[t]
\centering
\caption{Existing approaches to agentic state management \textnormal{(local: file and process state; remote: state of external services; L/R: consistency between local and remote state; C/E: consistency between agentic context and environment state; \tickPartial{} means partial support.)}}
\label{tab:prior}
\footnotesize
\renewcommand{\arraystretch}{0.85}
\newcommand{\priorclass}[3]{%
  \multirow{#1}{*}{\begin{tabular}{@{}l@{}}#2\\#3\end{tabular}}}
\begin{tabular}{l l cc cc ccc ccc}
\toprule
 & & \multicolumn{2}{c}{\textbf{Environment state}} & \multicolumn{2}{c}{\textbf{Consistency}} & \multicolumn{3}{c}{\textbf{Primitives}} & \multicolumn{3}{c}{\textbf{Stakeholders}} \\
\cmidrule(lr){3-4} \cmidrule(lr){5-6} \cmidrule(lr){7-9} \cmidrule(lr){10-12}
\textbf{Class} & \textbf{Approach}                    & \textbf{Local} & \textbf{Remote} & \textbf{L/R} & \textbf{C/E} & \textbf{Snapshot} & \textbf{Rollback} & \textbf{Fork} & \textbf{User} & \textbf{Harness} & \textbf{Agent} \\
\midrule
\priorclass{2}{Framework}{checkpoint} & LangChain~\cite{langchain}     & \tickNo        & \tickNo         & \tickNo      & \tickNo      & \tickNo           & \tickNo           & \tickNo       & \tickYes      & \tickYes         & \tickNo      \\
& OpenClaw~\cite{openclaw}       & \tickPartial   & \tickNo         & \tickNo      & \tickPartial & \tickPartial      & \tickPartial      & \tickNo       & \tickYes      & \tickPartial     & \tickNo      \\
\midrule
\priorclass{4}{Sandbox}{checkpoint} & Podman~\cite{podman}           & \tickYes       & \tickNo         & \tickNo      & \tickNo      & \tickYes          & \tickYes          & \tickYes      & \tickYes      & \tickYes         & \tickNo      \\
& E2B~\cite{e2b}                 & \tickYes       & \tickNo         & \tickNo      & \tickNo      & \tickYes          & \tickYes          & \tickYes      & \tickYes      & \tickYes         & \tickNo      \\
& CubeSandbox~\cite{cubesandbox} & \tickYes       & \tickNo         & \tickNo      & \tickNo      & \tickYes          & \tickYes          & \tickYes      & \tickYes      & \tickYes         & \tickNo      \\
& DeltaBox~\cite{deltabox}       & \tickYes       & \tickNo         & \tickNo      & \tickYes     & \tickYes          & \tickYes          & \tickYes      & \tickYes      & \tickYes         & \tickNo      \\
\midrule
\priorclass{3}{Workflow}{recovery} & SagaLLM~\cite{sagallm_pvldb_2025}         & \tickNo        & \tickYes        & \tickNo      & \tickYes     & \tickNo           & \tickPartial      & \tickNo       & \tickYes      & \tickYes         & \tickPartial \\
& LogAct~\cite{logact}           & \tickPartial   & \tickYes        & \tickPartial & \tickYes     & \tickNo           & \tickPartial      & \tickNo       & \tickYes      & \tickYes         & \tickYes     \\
& Cordon~\cite{cordon}           & \tickPartial   & \tickPartial    & \tickPartial & \tickNo      & \tickPartial      & \tickPartial      & \tickNo       & \tickYes      & \tickYes         & \tickNo      \\
\midrule
& \textbf{\sys}                        & \tickYes       & \tickYes        & \tickYes     & \tickYes     & \tickYes          & \tickYes          & \tickYes      & \tickYes      & \tickYes         & \tickYes     \\
\bottomrule
\end{tabular}
\end{table*}

We observe that supporting both effectiveness and safety requires \emph{state management} that enables exploration of alternative executions and recovery from unwanted changes: exploration allows alternative actions to be tried without their effects interfering, while recovery undoes unwanted effects. State management must cover local and remote state, and maintain consistency across these types of state and with the agent's context. Furthermore, recovery and exploration should be accessible to users, harnesses, and agents: a user may notice a mistake, a harness may organize a search, or an agent may decide to retry an action. Since each acts on the same environment, the guarantees must hold regardless of who initiates the operation. These state management capabilities must also be implemented efficiently to not degrade the performance of agent execution.

\subsection{Existing state management approaches}
\label{sec:background:challenege}

Prior approaches only support some of these requirements for state management. \T\ref{tab:prior} compares approaches along four dimensions: the \emph{environment state} columns describe whether local files, processes and remote service state are managed; the \emph{consistency} column refers to coordination between local and remote state~(L/R), and between agent context and environment state~(C/E); the \emph{primitives} column considers support for recording, restoring, and branching versions of the environment state; and the \emph{stakeholders} column identifies whether users, harness code, or agents can invoke state management.

\mypar{Framework checkpointing} Systems such as LangChain~\cite{langchain} and OpenClaw~\cite{openclaw} offer \emph{framework checkpoints} that preserve conversation history, intermediate results, and workflow progress. In terms of \emph{environment state}, this covers at most parts of the local state and excludes remote state. Their \emph{primitives} resume or revisit workflow steps, rather than restoring and branching the complete environment. Consequently, \emph{consistency} between context and environment is limited: rewinding a conversation does not undo a database update, and restoring workspace files does not restore the processes using them. For \emph{stakeholders}, these approaches support user- or harness-initiated recovery: orchestration code selects a saved workflow state and resumes execution from it. This does not itself provide agent-driven recovery, which would require exposing recovery choices to the LLM and interpreting its selection within the running agent loop.

\mypar{Sandbox checkpointing} Approaches such as Podman~\cite{podman}, E2B~\cite{e2b}, CubeSandbox~\cite{cubesandbox}, and DeltaBox~\cite{deltabox} provide \emph{sandbox checkpoints}. Their \emph{environment state} includes local files and processes, and their \emph{primitives} support snapshot, rollback, and branching through user and harness interfaces. Some also preserve context with local state, providing C/E \emph{consistency}, and recent systems reduce checkpointing costs for repeated exploration. However, remote service state remains outside sandboxes, so these guarantees do not extend to L/R consistency. A restored local report can thus disagree with database updates that remain committed. The fundamental limitation is state coverage: efficient local recovery does not undo remote effects. In terms of \emph{stakeholders}, the compared interfaces let users and harnesses request checkpoints and restores. Agent-selected recovery would require connecting checkpoint contents to task progress so that the agent can choose where to resume.

\mypar{Workflow recovery} Approaches such as SagaLLM~\cite{sagallm_pvldb_2025}, LogAct~\cite{logact}, and Cordon~\cite{cordon} address tool effects through corrective actions or delayed commitment. Their \emph{environment state} can include remote effects and selected local changes, but excludes running process state. In history-based approaches, histories and dependency records support C/E \emph{consistency} and guide recovery by harnesses or agents; L/R consistency remains limited to the effects covered by their recovery mechanisms. Their \emph{primitives} consequently have narrower semantics than restoring and branching arbitrary earlier environments. Compensation requires suitable undo operations and retained recovery information; staging supports abort before commitment, but cannot undo effects already released. Whether agent-driven primitives are supported depends on who controls recovery: approaches that expose execution histories to the LLM can ask it to diagnose failures and choose corrective actions;  transaction-based approaches carry out commit/abort decisions outside the agent's action-selection loop.

\tinyskip

\noindent
None of these prior approaches achieve all four dimensions. As summarized in \T\ref{tab:prior}, \sys combines these capabilities: it coordinates local and remote restoration, reflects the restored environment in agent context, and exposes the same state-management primitives to users, harnesses, and agents through a new unified state management abstraction.



\section{Agent Statepoints}
\label{sec:abstraction}

In this section, we introduce \emph{agent statepoints}, a new abstraction that unifies the management of an agent's environment state. We present the statepoint concept~(\S\ref{sec:abstraction:abstraction}), and describe the primitives that it enables~(\S\ref{sec:abstraction:implementation}).

\subsection{Concept}
\label{sec:abstraction:abstraction}

\begin{figure}[tb]
  \includegraphics[width=0.48\textwidth]{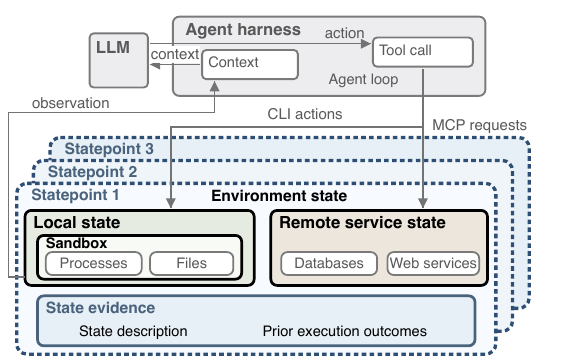}
  \caption{Overview of agent statepoints}
  \label{fig:statepoint}
\end{figure}

As \F\ref{fig:statepoint} shows, an \emph{agent statepoint} is a consistent, restorable point-in-time version of an agent's environment state, together with evidence that describes it. It provides a common object through which users, harnesses, and agents can reason about exploration and recovery. An agent action can change both local and remote state, so a statepoint groups them into a single version that can be recovered consistently.

A statepoint spans two state domains: the \emph{local state} comprises the files and running processes within the agent's sandbox. Statepoints require local tool execution to be confined to a sandbox: its boundary defines which files and processes must be captured and restored together, while isolating them from unrelated host state. The \emph{remote state} comprises external service state affected by the agent's supported operations, such as database updates through MCP requests.  Unlike the local state, the remote state is owned by the service provider instead of the agent, so the agent can observe and mutate this state only through the service API. The statepoint represents a recoverable version of this remote state without requiring the service to provide a checkpoint.

The agent statepoint must achieve \emph{cross-state consistency}: its local and remote state must reflect the agent-observable state at a single point in time. Otherwise, independently captured local and remote state could describe an environment that never existed. Treating them as one version makes consistency a requirement of the abstraction, rather than a responsibility of each caller.

Each statepoint also carries \emph{state evidence} that gives the environment version meaning: a description of its state and the outcomes of executions that previously proceeded from it. The evidence helps stakeholders (\eg the agent or user) select an appropriate statepoint to restore and avoid repeating executions that already failed. Evidence is associated metadata, distinct from both the captured environment state and the agent's LLM context. The evidence can be extended with a summary of what an execution attempted after that statepoint and what happened, such as completing the task or encountering an error. Adding this information leaves the recorded environment version unchanged. Keeping the evidence separate from the environment state allows knowledge of unsuccessful attempts to guide subsequent decisions even after returning to an earlier version.

\subsection{Primitives}
\label{sec:abstraction:implementation}

\begin{figure}[tb]
  \centering
  \includegraphics[width=0.410435\textwidth]{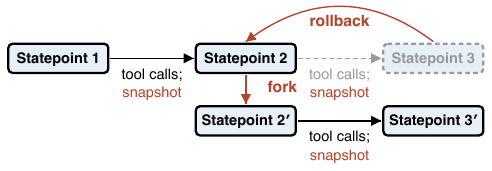}
  \caption{State management primitives for agent statepoints}
  \label{fig:primitive}
\end{figure}

The agent statepoint enables three state-management primitives (see~\F\ref{fig:primitive}): \emph{snapshot}, \emph{rollback}, and \emph{fork}, which meet the recovery and exploration requirements in~\S\ref{sec:background:state}. They operate on consistent environment versions and their associated state evidence, giving users, harnesses, and agents the same guarantees without requiring each caller to coordinate local and remote state.

\mypar{Snapshot} To return to an earlier execution point requires that the state was preserved consistently. \emph{Snapshot} captures both local and remote state and generates a new statepoint. It records the environment at a chosen boundary between tool calls as one restorable version, together with a state description. In~\F\ref{fig:primitive}, each transition from statepoint 1 to 2 and from 2 to 3 comprises one or more tool calls followed by an explicit snapshot; note that the abstraction does not require a snapshot after every call. Each statepoint covers local files and running processes as well as the affected remote service state, even when the service provides no checkpoint API.

\mypar{Rollback} Recovering from an error or abandoning an unsuccessful attempt requires undoing its effects across both state domains. \emph{Rollback} restores the environment state to a previously snapshotted statepoint.  In~\F\ref{fig:primitive}, returning from statepoint~3 to 2 abandons the later version shown in gray. For example, an agent records statepoint~2 after installing dependencies, then creates a file and registers it in a database. Rollback must undo both the file creation and the database update while preserving the dependencies. The state evidence helps callers choose where to return and identifies the restored environment. The agent's context is not rewound, and it retains knowledge of unsuccessful attempts~\cite{acrfence}.

\mypar{Fork} If the agent explores, the environment state of each attempt must remain isolated. \emph{Fork} creates a new environment branch from a previously captured statepoint. In~\F\ref{fig:primitive}, statepoint~$2'$ branches from statepoint~2; alternative tool calls followed by a snapshot produce statepoint~$3'$. Fork preserves the source branch, allowing it to continue independently. Local files and processes are isolated between branches; extending this isolation to remote state requires the service to support independent branches. The associated state evidence describes the common starting point and prior attempts, helping callers choose alternative actions.


\section{\sys Design}
\label{sec:design}

\sys is an agent runtime designed to realize the agent statepoint abstraction and state management primitives in~\S\ref{sec:abstraction}. In this section, we describe the overall design of \sys, and explain how it enables state management for agents. After that, \S\ref{sec:mechanism} and~\S\ref{sec:external} provide specific implementation details of how \sys manages local and remote state, respectively.

\subsection{Overview}
\label{sec:design:overview}

\begin{figure}[tb]
  \includegraphics[width=0.48\textwidth]{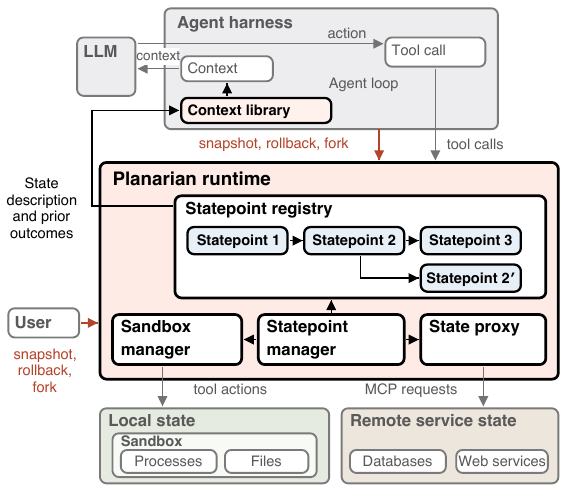}
  \caption{Design of \sys}
  \label{fig:design}
\end{figure}

\F\ref{fig:design}~shows \sys{}'s architectural components. During agentic execution, the state management functionality provided by \sys{} can be invoked by the user, the harness, or the agent itself. 
The harness routes tool execution through \sys and integrates with \sys via a context library.
\T\ref{tab:api} shows the state-management primitives \sys{} provides. A caller uses \texttt{snapshot} to record a statepoint, \texttt{rollback} to return the sandbox and remote state to that version, or the local \texttt{fork} operation to create another sandbox from its local state. The sandbox handle \texttt{sb} identifies the local execution environment, and \texttt{sp} identifies a statepoint, whose local and remote parts of state are \texttt{sp-local} and \texttt{sp-remote}. The \texttt{capture} and \texttt{compensate} hooks connect remote-service recovery to snapshot and rollback; the statepoint manager coordinates them with the sandbox operations.

\subsection{Local state management}

To enable state management, all tool invocations that read or mutate \emph{local} state (\eg processes, filesystems, etc.) must be run within a sandbox. This is already considered best practice for agentic systems in order to isolate the agent's execution from the host environment.

In \sys{}, a \emph{sandbox manager} creates and manages one or more sandboxes and dispatches the agent's tool calls to the appropriate sandbox. When a user, harness, or agent requests a snapshot, the sandbox manager captures the local state at a boundary between tool invocations. Callers can request a snapshot after each call or after several calls; the primitive does not prescribe a fixed frequency.

Delegating snapshot execution to the sandbox manager separates this choice from the mechanics of managing sandboxes and capturing their state. \sys uses Podman-based sandboxes~\cite{podman}, which combine incremental process checkpointing with zero-copy filesystem snapshots to reduce latency and copying overhead~(\S\ref{sec:mechanism}). Snapshot and fork can also overlap LLM inference~(\S\ref{sec:external:async}), which can hide latency~(\S\ref{sec:eval:exploration}).

\subsection{Remote state management}

For remote services without native versioning, \sys{} relies on \emph{compensating actions} that undo supported operations without requiring a service checkpoint. To achieve this, all tool calls that mutate remote state are routed through the \emph{state proxy}---a transparent interception layer between the agent harness and the target MCP server~(\S\ref{sec:external:proxy}). For each supported tool call, the state proxy transforms the original request into a compensable form and records the  information used to generate compensating actions. Requests that cannot be made compensable are rejected before execution.

When a statepoint is created, the state proxy records its current position in the sequence of MCP calls. On a \texttt{rollback(sp, sb, compensate)} request, it uses the recorded MCP call position to identify the subsequent actions and applies their compensating actions in reverse order. 

\begin{table}[t]
  \centering
  \caption{\sys{}'s state management APIs}
  \label{tab:api}
  \footnotesize
  \begin{tabular}{@{}>{\raggedright\arraybackslash}p{0.59\columnwidth}@{\hspace{3pt}}>{\raggedright\arraybackslash}p{\dimexpr0.41\columnwidth-3pt\relax}@{}}
    \toprule
    \textbf{Function} & \textbf{Description} \\
    \midrule
    \texttt{snapshot(sb,capture)} $\rightarrow$ \texttt{sp} & Records statepoint \texttt{sp} \\
    \texttt{rollback(sp,sb,compensate)} & Restores sandbox, service \\
    \texttt{fork(sp-local,sb)} $\rightarrow$ \texttt{sb'} & Forks sandbox from \texttt{sp-local} \\
    \midrule
    \texttt{capture(sb,service)} $\rightarrow$ \texttt{sp-remote} & Hook: records remote state \\
    \texttt{compensate(sp-remote,sb,service)} & Hook: restores \texttt{service} \\
    \bottomrule
  \end{tabular}
\end{table}

\subsection{State consistency}

To fully realize the statepoint abstraction (contribution~\#1), \sys must ensure that each statepoint is a consistent view of both the agent's local and remote state. This is handled by the \emph{statepoint manager} component, which coordinates between the sandbox manager and state proxy components. To create a new statepoint, it first freezes the sandbox, and takes the sandbox snapshot. After that, the MCP call position is captured from the state proxy. Once both parts are complete, the sandbox can be resumed.

During a snapshot, the statepoint manager holds an exclusive lock: it waits for any in-flight commands and MCP calls to finish before commencing the snapshot, and new calls are blocked until the snapshot completes~(\S\ref{sec:external:async}). The local and remote state are combined into a single statepoint, representing a point-in-time version of the agent's environment state, which is committed to the statepoint registry~(\S\ref{sec:external:registry}). 

Rolling back to a statepoint is also a two-step process: (i)~the statepoint manager applies the compensating actions through the state proxy, bringing the remote state to its logically equivalent point; and (ii)~it restores the local state of the sandbox from the snapshot.

The local \texttt{fork(sp-local, sb)} operation creates an isolated sandbox from a statepoint's local state, preserving the original sandbox. The local checkpointing mechanisms in \S\ref{sec:mechanism} support subsequent snapshots and rollbacks within each branch. Extending a branch to remote state requires service-side branching, \eg~from a database such as Neon~\cite{neon} or Dolt~\cite{dolt}. When a service supports compensation only, the forked sandbox cannot use that state proxy, preventing interference with the source branch~(\S\ref{sec:external:registry}).

\subsection{Agent context management}

To integrate the statepoint primitives into the agent's execution (contribution~\#2), the necessary information must be brought into the agent's context. In \sys{}, a \emph{context library} collects state evidence and supplies information about state-management operations to the harness. It makes evidence available to users, harnesses, and agents when choosing a statepoint for rollback, and supplies the selected state's description and prior outcomes after rollback. The harness retains responsibility for LLM invocation and context construction, including incorporating this information into the agent's context. It exposes several APIs to the harness, as shown in \T\ref{tab:context-api}:

\mypar{Snapshot creation} The context library gathers state from the harness, \eg~the progress of a task, the player information from a game, or the state of the sandbox. This constructs a mapping between a statepoint and its state, making each statepoint self-describing.
However, a static view of the agent's state at the time of statepoint creation would be insufficient. For example, the state at that point may appear to be suitable for restoration, but it might lead to a deadlock or an undesirable outcome if restored. Therefore, \sys also appends the execution outcomes that follow this statepoint, recording what previous executions did after this point.
The description and the outcomes form the state evidence~(\S\ref{sec:abstraction:abstraction}), which guides the choice of statepoint for rollback.
\F\ref{fig:statepoint_manifest} shows the data recorded for each statepoint.

\mypar{Statepoint selection} Before a caller chooses a statepoint for rollback, the context library exposes the \emph{state ledger}: the evidence associated with committed statepoints in the registry. Users and harnesses can inspect this evidence, and the harness can include it in the LLM's context so that the agent can choose a suitable statepoint.
Once rollback completes, the context library supplies the harness with the selected state's description and prior execution outcomes, together with notification that rollback has taken place. This information forms the \emph{restore context} returned by \texttt{RestoreContext}, which the harness appends to the context.
By design, this operation does not remove existing context, so as not to interfere with the harness's own context management, \eg~context compaction and memory. Without the restore context, the agent might make decisions based on outdated or inconsistent context, leading to repeated mistakes.

\begin{table}[t]
  \centering
  \caption{\sys{}'s context management APIs}
  \label{tab:context-api}
  \footnotesize
  \begin{tabular}{@{}lll@{}}
    \toprule
    \textbf{Action} & \textbf{Context API}         & \textbf{Description}       \\
    \midrule
    snapshot         & \texttt{StateRecord(...)}  & Records state description \\
                     & \texttt{AppendOutcome(...)}   & Appends outcomes \\
    \addlinespace
    selection    & \texttt{ConstructLedger(...)}   & Exposes state evidence \\
    rollback     & \texttt{RestoreContext(...)}    & Supplies rollback context \\
    \bottomrule
  \end{tabular}
\end{table}



\section{Local state management}
\label{sec:mechanism}

\sys is implemented as a harness-agnostic runtime that integrates with agent harnesses by routing tool calls through the \sys{} APIs. It uses Podman as the agent sandbox~(\S\ref{sec:mechanism:sandbox}), where the process tree is checkpointed using CRIU and the file system is snapshotted via ZFS~(\S\ref{sec:mechanism:snapshot}). \sys introduces a process checkpoint protocol that keeps CRIU's checkpoints incremental across restores~(\S\ref{sec:mechanism:criu}).

\subsection{Sandbox management}
\label{sec:mechanism:sandbox}
\sys{} provides a harness-agnostic sandbox manager that manages the lifecycle of the agent sandbox, including launching sandboxes, executing commands, and performing snapshot, rollback, and fork operations.
For agents that do not have a sandboxed execution environment, we integrate Podman as the sandbox backend, since it is a security-hardened OCI container engine with first-class checkpoint support. The sandbox manager launches a long-running Podman container with its built-in isolation primitives, \eg namespaces, cgroups, system call filtering, and access control, ensuring that an untrusted or erroneous agentic command cannot escape from the sandbox. \sys{} uses the Podman container as the boundary of the local state. Each container's root file system is mounted to a \sys{}-managed ZFS~\cite{zfs} dataset, which is a block-level copy-on-write (CoW) filesystem.

\subsection{Sandbox snapshot and the local state}
\label{sec:mechanism:snapshot}
The local state of a statepoint contains two parts: a process checkpoint and a file system snapshot. The container's processes form a process tree rooted at the container's PID~1. \sys{} uses CRIU~\cite{criu}, a process checkpoint tool, to checkpoint the agent container's process tree into an artifact. This enables \sys{} to restore all application processes running at the time of the snapshot.
\sys{} also snapshots the container's root file system, which on ZFS is a zero-copy, metadata-only operation. Together, they form a complete restore point of the sandbox. Using this mechanism, \sys can roll back the file system and the applications running inside it to a clean state, making the sandbox recoverable by preserving its state rather than reconstructing it.

\subsection{Continuous incremental checkpointing}
\label{sec:mechanism:criu}

To minimize the latency and storage cost of snapshots, \sys{} attempts to capture only the incremental changes since the last snapshot.
CRIU already supports incremental checkpoints, but with one notable limitation: the first checkpoint after a restore must always be a full (\ie non-incremental) checkpoint. There are two technical reasons for this:
(1) CRIU's incremental checkpointing relies on the kernel's soft-dirty bits. After restoring from a checkpoint, the process will have a new address space and new page table entries, thus losing incremental memory tracking (\ie all the soft-dirty bits will be set).
(2) CRIU has an internal safety mechanism based on checkpoint creation time. This mechanism is designed to prevent accidentally creating an incremental checkpoint based on a parent checkpoint from a different process, even if the process ID (PID) is the same (\eg through PID reuse).\footnote{For example, a process forked after a dump inherits the clean soft-dirty bits of the process it forked from, and it may reuse the PID of another checkpointed process that has since exited. The next dump only records the subsequently dirtied pages. When restoring from this incremental dump, CRIU would treat the exited process's checkpoint as the parent checkpoint of the forked process and restore the forked process with the correct dirty pages but an unrelated base memory state, resulting in incorrect memory.} 
Specifically, CRIU falls back to a full dump of a process whose start time is not earlier than the parent checkpoint's creation time, which prevents incremental checkpoints after a restore.

Since agentic workloads perform very frequent snapshot and restore operations during exploration, creating a full checkpoint after each restore becomes very costly.
Inspired by prior discussions in~\cite{criu-issue-401}, \sys proposes an incremental checkpoint protocol that requires no changes to the existing checkpoint/restore (C/R) stack (\ie~CRIU), but enables efficient incremental CRIU checkpointing across restores. The full protocol is shown in Algorithm~\ref{algo:checkpoint}.

Firstly, \sys{} uses the insight that a freshly restored process should not have any of the dirty bits set because it has been reconstructed directly from a valid checkpoint. \sys{} therefore uses CRIU's \texttt{pre-resume} hook (line~\ref{algo:checkpoint:restore}), which runs after the process tree has been restored but before execution resumes, to clear the soft-dirty bits of the restored processes (line~\ref{algo:checkpoint:reset}).

Secondly, since \sys{} can guarantee that the restored process is always a continuation of the checkpointed process, it can safely bypass CRIU's checkpoint creation time safety mechanism. Specifically, in the CRIU \texttt{pre-resume} hook (line~\ref{algo:checkpoint:epoch-compute}), \sys{} computes the epoch time $E$ as one tick after the latest start time of all restored threads. It then waits until the system uptime reaches or exceeds $E$ (line~\ref{algo:checkpoint:epoch}), making $E$ a boundary between the restored processes and any process created afterward. After the restore has completed, \sys modifies the creation time of the parent checkpoint to be the epoch time $E$ (\ie so that it appears the parent checkpoint was created \emph{after} the current process had started). This enables only the restored process to safely pass the checkpoint creation time check and create incremental checkpoints after the restore. If the restore was part of \sys{}'s fork primitive, \sys does not modify the shared restored image in place, as this could lead to two sandboxes writing the same image concurrently. Instead, \sys{} writes a per-sandbox copy of the restored image.

\begin{algorithm}[t]
\small
\caption{Continuous incremental checkpoint/restore}
\label{algo:checkpoint}
\begin{algorithmic}[1]
\makeatletter 
\newcommand{\BLOCK}[1]{\ALC@it #1\begin{ALC@g}}
\newcommand{\ENDBLOCK}[1]{\end{ALC@g}\ALC@it\algorithmicend\ #1}
\makeatother
\setlength{\itemsep}{1pt}
\setlength{\parsep}{0pt}
\setlength{\topsep}{0pt}
\REQUIRE sandbox $sb$, checkpoint chain $C=\langle c_0,\dots,c_n\rangle$, full-dump interval $k$, restore target $c_r \in C$, primitive~$smp$.
\medskip
\STATE \underline{\textsc{Checkpoint}$(sb, C, k)$}
\IF{$C = \emptyset \lor |C| \geq k$} \label{algo:checkpoint:dump-start}
    \STATE $c_{n+1} \gets \textsc{Dump}(sb,\ \texttt{track-mem})$;\ $C \gets \langle c_{n+1} \rangle$
\ELSE
    \STATE $c_{n+1} \gets \textsc{Dump}(sb,\ \texttt{track-mem},\ \texttt{parent}{=}c_n)$;\ $C \gets C \cdot c_{n+1}$
\ENDIF \label{algo:checkpoint:dump-end}
\medskip
\STATE \underline{\textsc{Restore}$(sb, C, c_r, smp)$}
\STATE $\textsc{Assemble}(\langle c_0,\dots,c_r\rangle)$ from $C$;\ \textbf{if} $smp = \mathit{rollback}$: $\textsc{Stop}(sb)$ \label{algo:checkpoint:restore-start}
\STATE $sb \gets \textsc{Restore}(sb,\ c_r,\ \texttt{hook}{=}\textsc{pre-resume})$ \label{algo:checkpoint:restore}
\BLOCK{\textbf{begin hook} \textsc{pre-resume}$(sb)$}
    \FORALL{$pid \in \textsc{Processes}(sb)$}
        \STATE $\textsc{Write}(\texttt{/proc/}pid\texttt{/clear\_refs},\ 4)$ \label{algo:checkpoint:reset}
    \ENDFOR
    \STATE $E \gets 1 + \max\{\, t.\mathit{start} : t \in \textsc{Threads}(sb) \,\}$ \label{algo:checkpoint:epoch-compute}
    \STATE \textbf{wait until} $\textsc{Uptime}() \geq E$;\ \textbf{return} $E$ \label{algo:checkpoint:epoch}
\ENDBLOCK{\textbf{hook}}
\IF{$smp = \mathit{rollback}$}
    \STATE $c_r.\mathit{dump\_time} \gets E$;\ $C \gets \langle c_0,\dots,c_r\rangle$ \label{algo:checkpoint:time}
\ELSIF{$smp = \mathit{fork}$}
    \STATE $c_r' \gets \textsc{Copy}(c_r)$
    \STATE $c_r'.\mathit{dump\_time} \gets E$;\ $C \gets \langle c_0,\dots,c_{r-1}, c_r'\rangle$
\ENDIF
\end{algorithmic}
\end{algorithm}



\section{Remote state management and consistency}
\label{sec:external}
Beyond the local sandbox state, \sys also captures the remote state~(\S\ref{sec:external:proxy}), managing these heterogeneous states consistently via statepoints using the state management primitives~(\S\ref{sec:external:registry}). \sys also provides asynchronous management of snapshot and fork, hiding the state management overhead under the LLM latency~(\S\ref{sec:external:async}).

\subsection{Remote state proxy}
\label{sec:external:proxy}

To make remote state restorable, \sys routes external tool calls through its state proxy component. This component provides similar functionality to the sandbox manager in that it is responsible for capturing and restoring the remote part of a statepoint.
When the agent interacts with an external service that supports versioning, \eg~a branchable database~\cite{dolt, neon}, the state proxy can simply use the service's existing functionality to create agent statepoints and roll back or fork from them. However, \sys{} also targets the more general case, where the external service does \emph{not} support versioning, and some of its operations are not compensable in their original form. To handle this, the state proxy first transforms each supported external tool call into a recoverable form and generates a compensating action, which can be used to \emph{undo} the original action.

\mypar{Proxy integration} To integrate the state proxy with an agent, the agent harness configures the state proxy as the MCP server in place of the target service's MCP server. When the agent issues an MCP call, the request is dispatched from the harness's MCP client to the state proxy. This proxy is both an MCP server (to the agent harness) and an MCP client (to the target MCP server), transparently intercepting all requests without changing the original MCP APIs.

\mypar{Request transformation} Upon receiving a request, the proxy first identifies whether the request can be compensated after transformation; unsupported requests that cannot be compensated are rejected before execution. The proxy then transforms the request into a compensable form. The rewrite and compensation logic is specific to the target MCP server. In \sys{}, we prototype the state proxy to support compensable SQL operations for the database MCP server. The protocol transforms a single SQL statement into a compensable form: a pre-image statement captures the necessary table state, the original statement mutates the database state, and a post-image statement records the keys affected by the original statement. These three statements are executed as a single SQL transaction, ensuring atomic execution. This makes the original statement compensable: deleted or modified rows can be restored by applying the values in the pre-image table, and added rows can be deleted according to the keys recorded in the post-image table.

\mypar{Compensation generation and enforcement} After rewriting, the state proxy generates the compensating action, forwards the rewritten request to the remote target MCP server, and records the compensating action in the undo log after successful execution. All proxy operations are executed under a proxy-level lock, ensuring that requests are serialized and recorded in a request log, so that subsequent compensating actions can be applied in the exact reverse order of their MCP requests in the log.

\begin{figure}[t]
  \begin{lstlisting}[style=tightjson, basicstyle=\ttfamily\scriptsize, backgroundcolor=\color{white}]
  // --- statepoint metadata --- //
  "statepoint-id": "...",
  "parent-statepoint-id": "...",
  "branch-id": "...",
  "timestamp": "...",
  "name": "...",
  "state-evidence": {
    "state-description": "...",
    "prior-outcomes": "...",
  },
  // --- statepoint artifacts ---  //
  "local-state": {
    "criu-checkpoint-path": "...",
    "zfs-snapshot": "...",
  },
  "remote-state": {
    "external-endpoint": "...",
    "mcp-call-position": "...",
    "hook-path": "...",
  },
  "status": "discarded| pending | committed"
\end{lstlisting}
\caption{\label{fig:statepoint_manifest}Example entry of an agent statepoint.}
\end{figure}

\begin{figure*}[t]
\begin{minipage}[b]{0.32\textwidth}
    \centering
    \includegraphics[width=\textwidth]{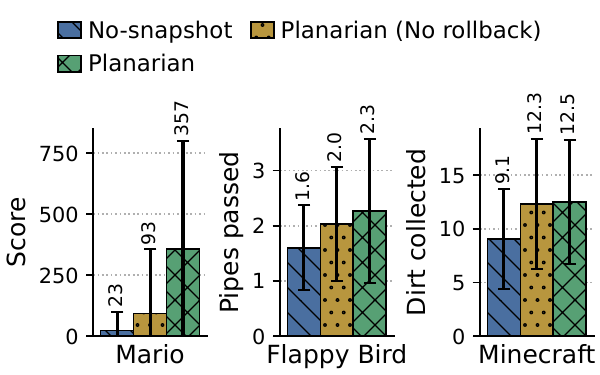}
    \caption{GameWorld, score}
    \label{fig:gameworld}
\end{minipage}
\hfill
\begin{minipage}[b]{0.32\textwidth}
    \centering
    \includegraphics[width=\textwidth]{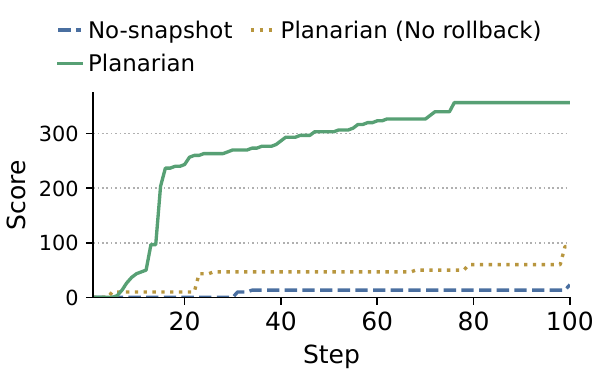}
    \caption{Mario, max score by step}
    \label{fig:gameworld_steps}
\end{minipage}
\hfill
\begin{minipage}[b]{0.32\textwidth}
    \centering
    \includegraphics[width=\textwidth]{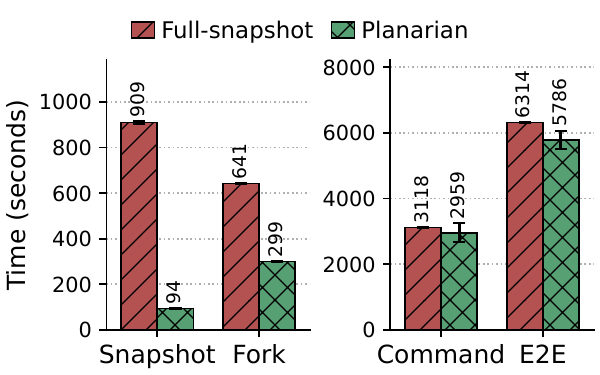}
    \caption{MCTS-style exploration}
    \label{fig:mcts}
\end{minipage}
\end{figure*}

\subsection{Statepoint manager and registry}
\label{sec:external:registry}

The statepoint registry manages a collection of the statepoints defined in \S\ref{sec:abstraction:abstraction}. Each registry entry (\F\ref{fig:statepoint_manifest}) includes all three artifacts in a single statepoint: CRIU's incremental process checkpoint, the zero-copy ZFS snapshot, and the remote state tracked by the state proxy.

To ensure each statepoint is a consistent slice of the agent's environment state that covers all heterogeneous resources that the agent mutates, \sys proposes a consistent snapshot and rollback/fork protocol on live, running sandboxes and services. At the start of \sys{}'s snapshot, the statepoint manager first freezes the container's processes via \texttt{cgroup} freeze.
It then takes the CRIU checkpoint and ZFS snapshot, and captures the remote state through the state proxy. The container's processes are unfrozen once all captures finish, ensuring that the sandbox state does not drift and thus stays consistent with the other artifacts. The statepoint manager only marks a statepoint as committed once all captures succeed; otherwise the statepoint manager marks it as pending and does not treat it as a valid statepoint. Rollback restores them as a single operation.

The above mechanism lets \sys maintain a stop-the-world invariant on the container across the entire snapshot window, including CRIU's internal process dump phase, where the processes remain frozen after the dump and resume only when \sys explicitly unfreezes the cgroup. This lets \sys freeze the container activity during the snapshot without stopping the Podman container, avoiding the cold-start overhead on every snapshot operation.

As \sys cannot quiesce a running external service that does not expose such APIs, \sys{}'s statepoint manager takes snapshots between tool calls. As each MCP server exposes a single service, one tool call typically mutates only one external service. Therefore, under the assumption that the external service is logically isolated by tenant (\ie the agent-readable and agent-mutable state of the external service is not affected by other tenants), \sys guarantees consistency across the sandbox and the external services.

On rollback, the statepoint manager first stops the container, then applies the compensating actions via the state proxy and restores the file system.
Once all other resources have been rolled back, CRIU starts its restore, making sure the container does not drift during the restore. Fork follows a similar pattern, but if the remote state is only compensable rather than forkable, \sys prevents the forked sandbox from using the state proxy, which would otherwise cause a consistency issue between exploration branches.

\subsection{Orchestrating snapshot and execution}
\label{sec:external:async}

Since the agent is waiting for the LLM's response between tool calls, the state management operations can be run in parallel with the LLM inference request. \sys is designed as a lightweight, daemonless runtime on top of Podman, CRIU, and ZFS, without requiring a dedicated runtime service. To make both snapshot and fork asynchronous, \sys launches background workers and coordinates them using a per-sandbox file lock. Tool calls, including command execution and MCP requests, acquire a shared lock, enabling concurrent tool execution. Snapshot and rollback either need the state to be frozen or mutate it, so they acquire an exclusive lock, waiting for in-flight tool calls to finish and preventing new tool calls from executing until they complete.
Since fork does not modify the source sandbox, it acquires a shared lock on the source sandbox and an exclusive lock on the forked sandbox, whose state it mutates. Snapshot and fork thus overlap with LLM inference while preserving statepoint consistency.


\section{Evaluation}
\label{sec:eval}

We evaluate \sys to answer the following questions: (i)~How does \sys perform under agentic exploration tasks~(\S\ref{sec:eval:exploration})? (ii)~What is \sys{}'s performance compared to existing state management approaches on system administration and coding tasks~(\S\ref{sec:eval:sandbox-rollback-e2e})? (iii)~What is \sys{}'s overhead compared to no state management~(\S\ref{sec:eval:co-rollback-e2e})? (iv)~What is the latency of \sys{}'s state management primitives~(\S\ref{sec:eval:microbench})?

\subsection{Experimental setup}
\label{sec:eval:setup}

\mypar{Testbed} We evaluate \sys on an Ubuntu 22.04 server with Intel Xeon Silver 4310 CPUs (24 cores, 48 threads), 128\unit{GB} of memory, and NVMe SSD storage.

\mypar{Implementation} We implement \sys in Go and Python. We use Podman~5.8.6 as the agent sandbox, ZFS~2.3.9 for sandbox storage, and CRIU~4.2.1 for process checkpointing. We also use the MCP toolbox for databases~\cite{toolbox} as an MCP server. Podman's container network is enabled when needed.

We integrate \sys with four agents: (i)~Coding---we use \texttt{mini-swe-agent}~\cite{minisweagent}, a lightweight software engineering agent designed for agentic coding tasks; (ii)~Personal assistant---we employ \texttt{nanobot}~\cite{nanobot}, a Python implementation of \texttt{openclaw}~\cite{openclaw} that automates everyday tasks; (iii)~Exploration harness---we use SWE-search~\cite{swesearch}, an MCTS-style tree-search harness for coding tasks; and (iv)~Gaming---we use the GameWorld~\cite{gameworld} harness and benchmark.

\mypar{Baselines} We compare \sys against two baselines: (i)~\vanilla, which executes agent tasks in a Podman container without state management. (ii)~\export, which runs \sys with Podman's native snapshot mechanism that exports a full snapshot of the container's process tree together with its file-system changes. Both baselines use an ext4 filesystem without CoW storage.

\mypar{Agent workloads} We evaluate \sys with four real-world agentic workloads: (i)~DevOps-Gym~\cite{devopsgym} consists of 704~DevOps tasks collected from over 30~projects; (ii)~Terminal-Bench 2.0~\cite{terminalbench} includes 89~tasks, spanning a wide range of categories from terminal environments; (iii)~Bird-interact-lite~\cite{birdinteract} is an interactive SQL-generation benchmark with 300~database tasks; and (iv)~GameWorld~\cite{gameworld} is a gaming benchmark with 34 popular games.

\begin{figure*}[t]
\centering
\includegraphics[width=0.44\textwidth]{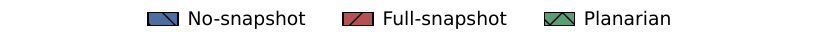}\\[2pt]
\begin{minipage}[b]{0.32\textwidth}
    \centering
    \includegraphics[width=\textwidth]{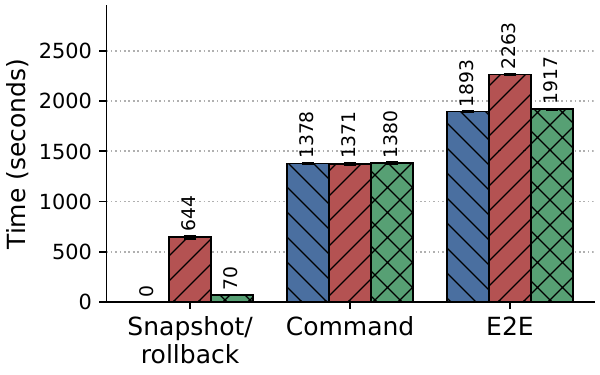}
    \caption{Terminal-Bench}
    \label{fig:e2e}
\end{minipage}
\hfill
\begin{minipage}[b]{0.32\textwidth}
    \centering
    \includegraphics[width=\textwidth]{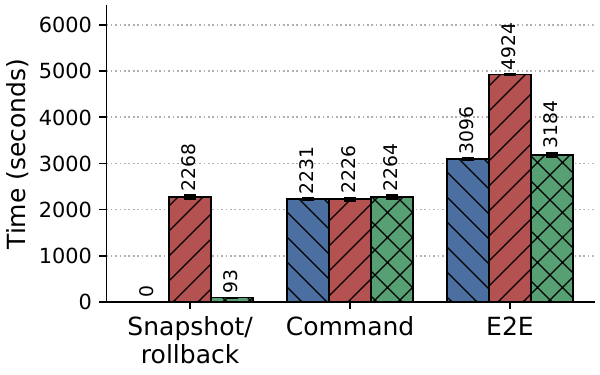}
    \caption{DevOps-Gym}
    \label{fig:e2e-devops}
\end{minipage}
\hfill
\begin{minipage}[b]{0.32\textwidth}
    \centering
    \includegraphics[width=\textwidth]{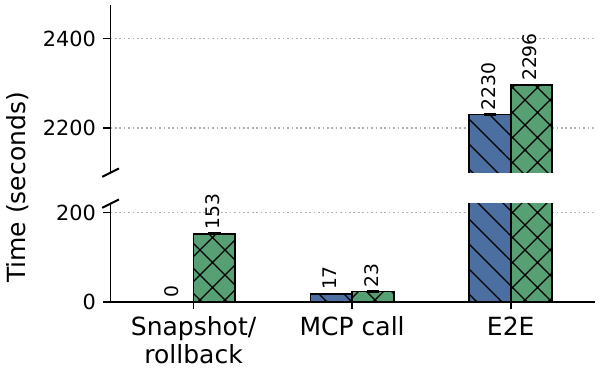}
    \caption{Bird-interact-lite}
    \label{fig:bird}
\end{minipage}
\end{figure*}

\begin{figure*}[t]
    \centering
    \includegraphics[width=0.72\textwidth]{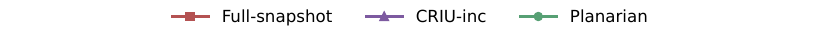}\\[2pt]
    \begin{subfigure}[b]{0.47\textwidth}
        \centering
        \includegraphics[width=0.48\linewidth]{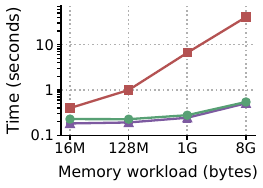}
        \hfill
        \includegraphics[width=0.48\linewidth]{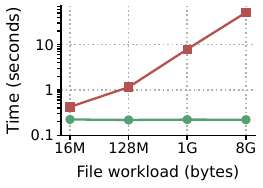}
        \caption{Snapshot before rollback/fork}
        \label{fig:micro_snap_before}
    \end{subfigure}
    \hfill
    \begin{subfigure}[b]{0.23\textwidth}
        \includegraphics[width=\textwidth]{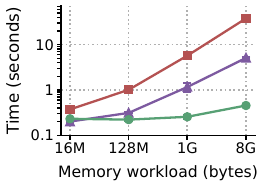}
        \caption{Snapshot after rollback}
        \label{fig:micro_snap_after_rollback}
    \end{subfigure}
    \hfill
    \begin{subfigure}[b]{0.23\textwidth}
        \includegraphics[width=\textwidth]{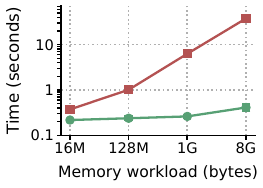}
        \caption{Snapshot after fork}
        \label{fig:micro_snap_after_fork}
    \end{subfigure}
    \caption{Snapshot cost}
    \label{fig:micro_snapshot}
\end{figure*}

\begin{figure*}[t]
    \centering
    \begin{minipage}[b]{0.47\textwidth}
        \centering
        \includegraphics[width=0.48\linewidth]{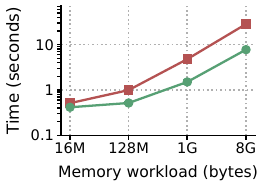}
        \hfill
        \includegraphics[width=0.48\linewidth]{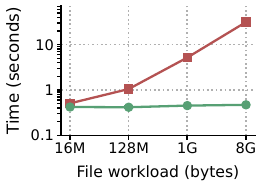}
        \caption{Rollback cost}
        \label{fig:micro_rollback}
    \end{minipage}
    \hfill
    \begin{minipage}[b]{0.47\textwidth}
        \centering
        \includegraphics[width=0.48\linewidth]{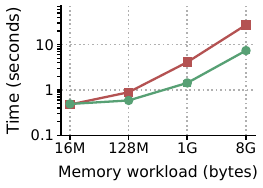}
        \hfill
        \includegraphics[width=0.48\linewidth]{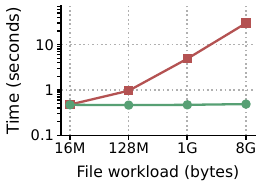}
        \caption{Fork cost}
        \label{fig:micro_fork}
    \end{minipage}
\end{figure*}

\subsection{Agentic exploration with state management}
\label{sec:eval:exploration}

\mypar{LLM-driven exploration} First, we measure how the LLM can proactively interact with \sys to explore the execution space and safely try different rollouts to find better execution plans in fewer steps. For this, we integrate \sys with GameWorld, enabling the gaming agent to roll back to a previous statepoint. When the game reaches a terminal state in a round, the LLM reviews all available statepoints in the state ledger and either rolls back to a previous statepoint or resumes from the reset game state; after a rollback, the restore context is appended to the following prompts. The state ledger and restore context are constructed from the live game state of the GameWorld harness, with player-unobservable state filtered out.\footnote{We define player-unobservable state as state that cannot be reconstructed from historical game screenshots.} Statepoints are captured between game plays, and only non-terminal states are available for rollback.

We evaluate \sys on the GameWorld~\cite{gameworld} benchmark with three diverse games: Mario, Flappy Bird, and Minecraft. For each, we use the difficulty level at which the agent makes the most absolute progress toward the goal within the step limit. Note that the GameWorld-provided Minecraft game is a simplified item-collection simulator: to approximate the real game's difficulty, we add real hazards to this simulator, including creepers, lava damage, etc.

As shown in \F\ref{fig:gameworld}, \sys enables the LLM to find a better strategy within 100~steps compared to no state management: the agent achieves 15$\times$, 1.4$\times$, and 1.4$\times$ the \vanilla scores in Mario, Flappy Bird, and Minecraft, respectively. When the game enters a bad state (\eg a reset due to death), the LLM uses \sys to roll back to a previous good state, explores alternative plans (as guided by the restore context), and achieves a higher score (see~\F\ref{fig:gameworld_steps}).

Note that snapshots may affect the timing of frames and may change the outcomes. To rule out this effect, we add a \textsc{\sys (No rollback)} baseline, which uses \sys to snapshot after each game step but without exposing the rollback primitive. This variant achieves a higher score than \vanilla, but still lower than full \sys.

\mypar{Harness-driven exploration} We integrate \sys with SWE-search~\cite{swesearch}, an MCTS-style tree search harness that explores the execution space of software-engineering tasks using Git-based versioning and branching. Since \sys supports snapshot/fork over the full sandbox state, we extend this harness to MCTS-style tree search on the Terminal-Bench tasks. Nodes in the search tree are snapshotted and forked on demand, overlapping these operations with LLM invocations. We use \texttt{GPT-5.6-luna} (medium reasoning) to generate MCTS-style tree-search traces, and replay the full traces, including the recorded LLM latency, command execution, and snapshot/fork operations on both \sys and \export. We compare \sys with \export{}: \vanilla is not included, because it lacks the fork primitive needed for MCTS-style search.

\F\ref{fig:mcts} shows the overhead of MCTS-style tree search with \sys and \export on nine system-administration tasks from Terminal-Bench, broken down into fork, snapshot, command execution, and end-to-end~(E2E) time; the latter partially overlaps with snapshot and fork. \export{}'s primitives are significantly more expensive: its snapshot overhead is 10$\times$ that of \sys, because \sys checkpoints incrementally after a process restore, and its fork overhead is 2$\times$ that of \sys. As a result, \export is 9\% slower than \sys in end-to-end time.

The end-to-end gap is smaller than the primitive gap, because \export also benefits from overlap: each node in SWE-search issues two LLM calls, and the resulting LLM latency is long enough to hide most of \export{}'s overhead. With more frequent snapshot/fork operations or with lower LLM latency, less of that overhead can be hidden, and the end-to-end gap grows toward the performance gap of the snapshot/fork primitives.

\subsection{Safe execution with state management}
\label{sec:eval:sandbox-rollback-e2e}

We next evaluate \sys{}'s performance when used for user-driven rollback. We first use the \sys{}-integrated agent with \texttt{GPT-5.6-luna} (medium reasoning) to generate tool-calling traces, then replay these traces on \sys and the two baselines, because LLM outputs are non-deterministic.

\mypar{Terminal} In this experiment, the mini-swe-agent runs on \sys and executes tasks from Terminal-Bench (nine tasks from the system-administration category). \sys snapshots the agent sandbox after each tool call, aligning with the granularity of human approval. At the end of execution, both \sys and \export roll back to a previous snapshot.

\F\ref{fig:e2e} shows that, compared with \vanilla, \sys adds a 1\% overhead, which is significantly faster than \export (20\%), because \export{}'s snapshots are more heavyweight. Compared with \export, \sys reduces raw snapshot/rollback times by 89\%. Most Terminal-Bench tasks are not filesystem heavy, so copy-on-write does not add significant overhead.

\mypar{DevOps} We integrate mini-swe-agent with \sys to evaluate performance on DevOps tasks. We use DevOps-Gym's 16~implementation tasks in the build/configuration category. \F\ref{fig:e2e-devops} shows that \sys requires less than 3\% extra task completion time compared to  \vanilla; \export adds a 59\% overhead. Since \export{}'s sandbox snapshot/rollback time is $\sim$24$\times$ that of \sys, it becomes challenging to hide the snapshot time behind the LLM inference time. 

\subsection{Overhead of consistent state management}
\label{sec:eval:co-rollback-e2e}

To evaluate \sys{}'s overhead when both local and remote state are managed consistently, we execute 50~database management tasks from the Bird-interact-lite~\cite{birdinteract} benchmark, in which an agent mutates a PostgreSQL database via MCP~\cite{toolbox}. For \sys, the state proxy generates compensating actions and rewrites SQL requests. We use \texttt{GPT-5.6-luna} (medium reasoning) to generate traces for MCP requests. As in \S\ref{sec:eval:sandbox-rollback-e2e}, we then replay the requests on both \vanilla and \sys, with \sys snapshotting after each request and rolling back the sandbox and database.

\F\ref{fig:bird} shows that, while managing local and remote state consistently, \sys adds a 3\% overhead to the end-to-end time. This overhead comes from three sources: (i)~when interacting with the MCP database server, the harness sometimes sends several requests instead of interleaving them with LLM inference, preventing snapshot time from being hidden; (ii)~\sys incurs extra rollback operations for both the sandbox and database state; and (iii)~to make MCP requests revertible, the state proxy adds SQL operations: \sys takes 1.3$\times$ the MCP round-trip time compared to the \vanilla baseline. Since agentic database operations and MCP requests have low latency, this request amplification only increases end-to-end time by less than 0.3\%.

\subsection{State management primitives}
\label{sec:eval:microbench}

We also evaluate the overhead of \sys{}'s state management primitives under different microbenchmarks. We use a workload that dirties memory or the file system. We take two successive snapshots and measure the overhead of the second, which is incremental when supported. We then measure the overhead of rolling back to, or forking from, this second statepoint. For the memory workload, we also measure the latency of the first snapshot taken after the rollback and after the fork. We compare \sys with \export and with \textsc{CRIU-inc}, which wires CRIU's native incremental checkpointing through Podman.

\F\ref{fig:micro_snapshot} shows the overhead under these workloads. With a large working set~(8\unit{GB}), \sys{}'s incremental snapshot achieves 74$\times$ and 231$\times$ speedups over \export on the memory and file system workloads, respectively, because \export always provides a full dump. \sys{}'s continuous incremental checkpoint protocol is also faster than CRIU's native incremental checkpointing for the first snapshot after rollback: on the 8\unit{GB} memory workload, \sys provides an 11$\times$ speedup over \textsc{CRIU-inc}, because \sys continues to checkpoint incrementally after a restore.

We also measure the overhead of the rollback and fork primitives on the incremental snapshot (see~Figs.~\ref{fig:micro_rollback} and~\ref{fig:micro_fork}). For rollback, \sys achieves 4$\times$ and 67$\times$ speedups over \export on the 8\unit{GB} memory and file system workloads, respectively. Restoring the filesystem is faster than restoring memory, due to the cost of reconstructing a large process. Fork shows a similar pattern: for a sandbox with the same memory and file system state, \sys exhibits 4$\times$ and 61$\times$ speedups over the same baseline. We conclude that, with the incremental checkpoint protocol and CoW storage, \sys provides efficient state management primitives.


\section{Related Work}
\label{sec:rel_work}

\mypar{State recovery for agents} Existing systems version the state an agent mutates, either to recover the system from the agent's mistakes or to make the agent's state portable across machines. AgentFS~\cite{agentfs} is a SQLite-backed overlay file system that makes the agent's file system state checkpointable and portable. It covers file system state only, and a checkpoint copies the entire writable overlay layer. Daytona~\cite{daytona} and E2B~\cite{e2b} are cloud sandbox providers that offer persistent sandbox snapshots from which new sandboxes are created. They focus on the sandbox state itself, leaving remote state out of scope. Similarly, Agent Sandbox~\cite{agent-sandbox} supports persistent, stateful execution of Kubernetes (k8s) pods, so that idle agent sandboxes can be paused. k8s is designed to host remotely deployed microservices, whereas \sys targets the user's local deployment; \sys can integrate with Agent Sandbox to make such remotely deployed sandboxes recoverable.

Other work makes external database state recoverable. Dolt~\cite{dolt} provides Git-like operations on database state, enabling users to revert, branch, and merge the agent-mutated database. Similarly, Neon~\cite{neon} provides a CoW branching primitive for PostgreSQL. Both branchable database services can be integrated with \sys through its hooks, enabling consistent rollback and branching across both local and remote state.

\mypar{Agentic explorations} SWE-search~\cite{swesearch} is an MCTS-style search that executes, evaluates, and expands high-reward nodes, enabling search over Git-based software engineering tasks. 
\sys extends this beyond Git, enabling tree search on arbitrary terminal tasks. ExACT~\cite{exact} provides LLM-driven MCTS exploration for GUI tasks and relies on backtracking in the web environment. \sys can integrate with ExACT, enabling backtracking even when the workload does not expose this capability.

\mypar{Snapshot/restore systems} Various snapshot and restore systems support capturing a live system state and rollback when needed. Checkpoint/Restore In Userspace (CRIU)~\cite{criu} is a userspace process checkpointing tool that captures the state of a process tree, including memory pages, active sockets, and file descriptors, into a restorable image. As an alternative, DMTCP~\cite{dmtcp_ipdps_2009} is a process checkpointing library that dumps process state by preloading its library into the application at launch. We use CRIU for process checkpointing in \sys since it is widely integrated with OCI runtimes.
Snapshots are also natively supported by many file systems and device mappers. The Zettabyte File System (ZFS)~\cite{zfs} is a CoW file system that pools the available storage and manages the pool state as a Merkle tree. Since it provides block-level CoW, \sys integrates ZFS for fast file system snapshots. Other CoW snapshot mechanisms for persistent state include Btrfs~\cite{btrfs}, a Linux-native file system that provides CoW over a B-tree, and dm-snapshot~\cite{dm-snapshot}, a kernel device mapper that provides fast snapshots of a block device.


\section{Conclusions}
\label{sec:concl}
We presented \sys{}, an agent runtime that manages local and remote environment state through \emph{agent statepoints}: consistent, restorable point-in-time versions of the environment. Three primitives---\emph{snapshot}, \emph{rollback}, and \emph{fork}---give users, harnesses, and agents a unified interface for recovery and exploration, while the context library keeps agents informed of restored state and past outcomes. Our evaluation shows that \sys improves task quality by enabling agents to explore alternatives and lets users recover from erroneous agent actions with little overhead.

\bibliographystyle{ACM-Reference-Format}
\bibliography{agent-runtime}

\end{document}